\documentclass{article}
\usepackage{spconf,amsmath,graphicx,url}

\title{Deep Complex U-Net with Conformer for Audio-Visual Speech Enhancement}
\name{%
  \begin{tabular}[t]{c}
    Shafique Ahmed\textsuperscript{1,2}, Chia-Wei Chen\textsuperscript{3}, Wenze Ren\textsuperscript{3}, Chin-Jou Li\textsuperscript{3}, Ernie Chu\textsuperscript{1}, Jun-Cheng Chen\textsuperscript{1},\\
    \textit{Amir Hussain\textsuperscript{4}, Hsin-Min Wang\textsuperscript{1}, Yu Tsao\textsuperscript{1}, and Jen-Cheng Hou\textsuperscript{1}}
  \end{tabular}}

\address{\small{\textsuperscript{1}Academia Sinica \textsuperscript{2}National Tsing Hua University,  \textsuperscript{3}National Taiwan University \textsuperscript{4}Edinburgh Napier University}}

\begin{document}
%\ninept
%
\maketitle
\begin{abstract}
Recent studies have increasingly acknowledged the advantages of incorporating visual data into speech enhancement (SE) systems. In this paper, we introduce a novel audio-visual SE approach, termed DCUC-Net (deep complex U-Net with conformer network). The proposed DCUC-Net leverages complex domain features and a stack of conformer blocks. The encoder and decoder of DCUC-Net are designed using a complex U-Net-based framework. The audio and visual signals are processed using a complex encoder and a ResNet-18 model, respectively. These processed signals are then fused using the conformer blocks and transformed into enhanced speech waveforms via a complex decoder. The conformer blocks consist of a combination of self-attention mechanisms and convolutional operations, enabling DCUC-Net to effectively capture both global and local audio-visual dependencies. Our experimental results demonstrate the effectiveness of DCUC-Net, as it outperforms the baseline model from the COG-MHEAR AVSE Challenge 2023 by a notable margin of 0.14 in terms of PESQ. Additionally, the proposed DCUC-Net performs comparably to a state-of-the-art model and outperforms all other compared models on the Taiwan Mandarin speech with video (TMSV) dataset.
\end{abstract}

\section{INTRODUCTION}
\label{sec:intro}

Effective communication through spoken language is the cornerstone of human interaction. In various fields such as telecommunications, voice assistants, hearing aids, and video conferencing, ensuring the quality and intelligibility of speech is crucial. However, the challenge lies in the ability to consistently achieve good quality and intelligibility, especially in adverse acoustic environments characterized by background noise, reverberation, or limited audio quality. In the pursuit of addressing these challenges and enhancing speech quality and intelligibility, speech enhancement (SE) has emerged as a critical area of research and development~\cite{loizou2013speech}. The field of SE has witnessed significant advancements through the integration of deep learning methods. While deep learning-based SE techniques~\cite{lu2013speech, 6853860} have demonstrated remarkable success by primarily focusing on audio signals alone, it is notable that incorporating the visual modality can bring substantial benefits to improving the performance of SE systems in challenging acoustic environments~\cite{8323326, 9156852, 10.1109/TASLP.2021.3066303}.

Depending on the input type, audio-only SE methods can be roughly divided into two categories: time-domain methods and time-frequency (TF) domain methods. Conventional TF domain methods typically rely on amplitude spectral features; yet, studies have shown that SE performance can be limited because phase information is not adequately considered ~\cite{6853860}. To address this issue, approaches that employ complex-valued features, such as complex spectral mapping (CSM)~\cite{8682834} and complex ratio masking (CRM)~\cite{williamson2017time}, as SE input have recently been proposed. Many CSM and CRM techniques are built upon real-valued neural networks, while others employ complex-valued neural networks to process the complex input. Deep Complex U-NET (DCUNET)~\cite{choi2018phaseaware} and Deep Complex Convolution Recurrent Network (DCCRN)~\cite{hu2020dccrn} are notable complex-valued neural networks for the SE task. In this study, we also use a complex-valued neural network to process audio data. 

The core idea behind audio-visual speech enhancement (AVSE) is to integrate visual input as supplementary data into an audio-only SE system, aiming to improve SE performance with the help of supplementary information. Several earlier studies have demonstrated the efficacy of incorporating visual input to enhance the performance of SE systems~\cite{8323326,DBLP:journals/corr/abs-1804-04121,chuang2022improved}. Most previous AVSE systems focused primarily on processing audio in the TF domain~\cite{10193049,10.1109/TASLP.2021.3066303}. Some previous studies performed audio-visual speech separation tasks in the time domain~\cite{9746866}. Recently, self-supervised learning (SSL) embeddings have been used to improve the performance of AVSE. For example, Richard et al.~\cite{lai2023audiovisual} introduced the SSL-AVSE method, which integrates visual cues with audio signals. These combined audio-visual data are fed into a Transformer-based SSL AV-HuBERT model to extract features, which are subsequently processed using a BLSTM-based SE model. 

In this paper, we propose a novel AVSE approach, termed DCUC-Net (deep complex U-Net with conformer network). DCUC-Net employs a deep complex U-Net architecture to incorporate phase information into AVSE. The audio stream is processed by a complex encoder to create a complex representation of the audio data and then combined with visual features. To further enhance the combined audio-visual features, DCUC-Net integrates conformer blocks into the deep complex U-Net architecture. These conformer blocks enable DCUC-Net to effectively capture long-range dependencies, both local and global, and fine-grained contextual information between audio and visual modalities. The resulting features are then decoded by a complex decoder to estimate a complex mask. By multiplying the complex spectra from the noisy input with the estimated complex mask, we can obtain enhanced complex spectra, which are then converted to enhanced waveforms as the final output. We evaluate the performance of DCUC-Net through experiments on the dataset used in the COG-MHEAR
AVSE Challenge 2023, which is built on the LRS3 dataset~\cite{afouras2018lrs3ted}, and the Taiwan Mandarin speech with video (TMSV) dataset. The results demonstrate the efficacy of DCUC-Net, as it surpasses existing baselines and demonstrates robust denoising capabilities.

\begin{figure}[htb]

\begin{minipage}[b]{1.0\linewidth}
  \centering
  \centerline{\includegraphics[width=8.5cm]{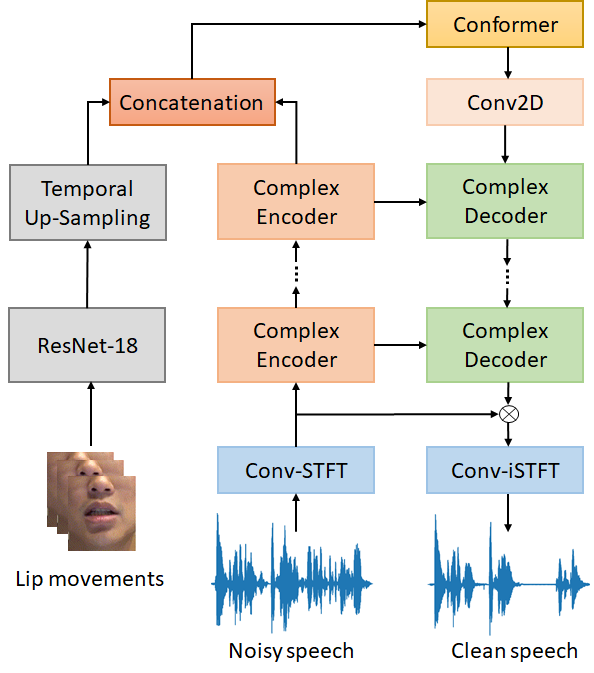}}
%  \vspace{2.0cm}
\end{minipage}
\caption{Architecture of the proposed DCUC-Net.}

\label{fig:arc}
\end{figure}

\section{The Proposed DCUC-Net AVSE System}
\label{sec:method}

The proposed DCUC-Net aims to obtain enhanced speech signals with improved intelligibility and quality via the integration of audio and visual information. Figure~\ref{fig:arc} shows the overall architecture of DCUC-Net. In DCUC-Net, we utilize ResNet-18 to process lip movement cues, thereby preparing the visual information. Additionally, we employ a complex encoder to transform the audio input into complex-valued features. The complex encoder of DCUC-Net comprises five Conv2d blocks, which are designed to extract encoded features from the audio. These extracted audio features are then combined with visual information, and we leverage conformer blocks~\cite{Conformer} to effectively process this combined audio-visual information. During decoding, the symmetric complex-valued encoder-decoder structure is considered and used to reconstruct the lower-dimensional representation to the original size of the input, facilitating the restoration of the enhanced speech waveforms.
The complex encoder and decoder block in DCUC-Net is constructed based on the implementation details described in~\cite{trabelsi2018deep}. It comprises three main components: complex Conv2d~\cite{hu2020dccrn}, complex batch normalization~\cite{trabelsi2018deep}, and real-valued PReLU~\cite{C4}. The complex Conv2d operation enables the manipulation of complex-valued features. In this operation, the complex-valued convolutional filter, denoted as $W$, is defined as \(W = W_r + jW_i\), where \(W_r\) represents the real part, and \(W_i\) represents the imaginary part. Similarly, the input complex spectra, denoted as $X$,  is defined as \(X = X_r + jX_i\), where \(X_r\) and \(X_i\) denote the real and imaginary parts, respectively. The complex convolution operation, denoted as \(X \bigotimes W\), is performed to obtain the complex output.
The output feature of each complex layer, represented as \(F_{out}\), is calculated as:

\begin{equation}
\begin{aligned}
F_{out}= (X_r \times W_r - X_i \times W_i) + j(X_r \times W_i + X_i \times W_r). 
\end{aligned}
\end{equation}
Equation (1) within DCUC-Net enables the manipulation of both the real and imaginary parts of the complex-valued features, facilitating effective processing of the audio part.

Expanding on this capability, we integrate the extracted visual features into DCUC-Net by concatenating them with the real values from the complex audio feature. However, this concatenation process introduces a sequential misalignment between the visual and audio features. To address this discrepancy, we employ a temporal upsampling technique to align the temporal sequences of the visual features with the temporal resolution of the audio features. This alignment process ensures proper fusion of the two modalities. Subsequently, the aligned audio and visual features are input into the Conformer blocks.
% Expanding on this capability, we integrate the extracted visual features into DCUC-Net by concatenating them with the complex encoder features. For the concatenation process, we have only considered real values from the complex audio feature. However, we encounter a sequential difference between the visual and audio features during the concatenation process. To address this discrepancy, we employ a temporal upsampling technique that aligns the temporal sequences of the visual features with the temporal resolution of the audio features. This alignment process ensures proper fusion of the two modalities. Subsequently, the aligned audio and visual features are fed into the conformer blocks.

The Conformer block comprises two Feed-Forward (FFN) modules positioned around the Multi-Headed Self-Attention (MHSA) module and the Convolution (Conv) module. In contrast to the original Transformer block, which incorporates a single feed-forward layer, the Conformer block features two half-step feed-forward layers: one preceding the attention layer and one succeeding it. To put it simply, when the $i$ Conformer block receives an input, denoted as $x_i$ the resulting output $y_i$ is determined through the following mathematical expression:

\begin{equation}
\begin{aligned}
    \hat{x}_i &= x_i + \frac{1}{2} \text{FFN}(x_i) \\
    \check{x}_i &= \hat{x}_i + \text{MHSA}(\hat{x}_i) \\
    \bar{x}_i &= \check{x}_i + \text{Conv}(\check{x}_i) \\
    y_i &= \text{Layernorm}(\bar{x}_i + \frac{1}{2} \text{FFN}(\bar{x}_i))
\end{aligned}
\end{equation}

The conformer blocks play a critical role in capturing both global and local audio-visual dependencies within our DCUC-Net. These blocks facilitate the effective combination of the audio and visual modalities by leveraging self-attention mechanisms and convolutional operations. The self-attention mechanisms establish relationships between different parts of the audio and visual features, enabling the model to capture long-range dependencies and effectively leverage global audio-visual cues. Simultaneously, the convolutional operations extract local contextual information, ensuring the model's robustness in capturing fine-grained audio-visual details.

With the incorporation of the audio and visual information, the output of the conformer blocks is passed to the complex decoder component in DCUC-Net. The role of this decoder is to estimate the complex ratio mask and obtain enhanced complex spectra by multiplying them with the noisy complex spectra. Subsequently, the complex spectra are transformed into an enhanced speech waveform by applying the Convolutional inverse Short-Time Fourier Transform (ConviSTFT). It's worth noting that the decoder benefits from a skip-connection mechanism, which provides access to encode features via the U-Net architecture.

% \begin{figure}[htb]

% \begin{minipage}[b]{1.0\linewidth}
%   \centering
%   \centerline{\includegraphics[width=5.0cm]{intro_3.png}}
% %  \vspace{2.0cm}
% \end{minipage}
% \caption{The Conformer block.}

% \label{fig:conformer}
% %
% \end{figure}

\section{EXPERIMENTS}
\label{sec:experiments}

\subsection{Datasets}
We evaluated the proposed DCUC-Net on two datasets: the COG-MHEAR Audio-Visual Speech Enhancement Challenge 2023~\cite{challenge2023} dataset and the Taiwan Mandarin speech with video (TMSV) dataset. The COG-MHEAR dataset is built on the LRS3 dataset and contains two types of interference sources: speech and noise. The speech interference source comes from the LRS3 dataset, while the noise interference source comes from three different sources: the 1st Clarity Enhancement Challenge, DEMAND~\cite{demand}, and Deep Noise Suppression Challenge (DNS)~\cite{reddy21_interspeech} 2nd version.
In the training set, interference sources come from 405 competing speakers and 7,346 noise files covering 15 noise categories. There are a total of 34,524 scenes with 605 target speakers. On the other hand, For the test set, we utilized the development set provided by the challenge, in which interference sources are selected from 30 competing speakers and 1,825 noise files in the same 15 noise categories mentioned above. The test set contains 3,306 scenes with 85 target speakers. The TMSV dataset comprises video recordings from 18 native Mandarin speakers (13 males and 5 females). Each speaker recorded 320 Mandarin sentences, with each sentence consisting of 10 Chinese characters. The duration of each utterance ranges from approximately 2 to 4 seconds. To ensure methodological congruence and experiment reproducibility, we followed the procedures detailed in~\cite{chuang2022improved} for the introduction of noise interference into the dataset and train-test split.

% The TMSV dataset contains video recordings of 18 native Mandarin speakers (13 males and 5 females). Each speaker recorded 320 Mandarin sentences, each sentence consisting of 10 Chinese characters. The duration of each utterance is approximately 2 to 4 seconds. For the TMSV dataset train and test split, as well as the noise interference, we followed the procedure details provided in~\cite{chuang2022improved} to ensure methodological congruence and reproducibility of our experiments.

\subsection{Experimental Setup}
\label{ssec:setup}
DCUC-Net preprocesses the input audio using a Convolutional Short-Time Fourier Transform (ConvSTFT) with a window length of 400 samples, a window increment of 160 samples, and an FFT length of 512 samples. The model encoder consists of 5 complex Conv2D blocks for audio processing, ResNet-18 for visual feature extraction (followed by temporal upsampling), 2 conformer blocks for capturing audio-visual dependencies, and 5 decoder blocks for estimating the complex mask. The enhanced speech signal is obtained by multiplying the estimated mask with the complex spectrogram and applying the Convolutional inverse Short-Time Fourier Transform (Conv-iSTFT).

To train DCUC-Net, we employed the Scale-Invariant Signal-to-Noise Ratio (SI-SNR)~\cite{Luo_2019} as the loss function, which is defined as: 
\begin{align}
\text{SI-SNR} &= 10  \log_{10} \left( \frac{{\|s_{\text{target}}\|^2}}{{\|e_{\text{noise}}\|^2}} \right), \label{eq:si-snr}
%\notag
\end{align}
where
\text{$\|s_{\text{target}}\|^2$} is the power of the target speech signal, and
\text{$\|e_{\text{noise}}\|^2$} represents the power of the estimated noise. The estimated noise is obtained by subtracting the target speech signal from the estimated speech signal.
SI-SNR is a commonly used evaluation metric that provides a more robust measure of performance than mean square error (MSE).

\subsection{Evaluation Results on the COG-MHEAR Dataset}
\label{ssec:COG-MHEAR results}

We evaluated three types of speech, namely noisy speech, enhanced speech generated by baseline models from the COG-MHEAR AVSE Challenge in 2022 and 2023, and enhanced speech generated by DCUC-Net, with two standard metrics: Perceptual Evaluation of Speech Quality (PESQ) and Short-Time Objective Intelligibility (STOI). Several noteworthy observations can be drawn from the results in Table 1. First, compared with noisy speech, the enhanced speech output of all AVSE models exhibits superior quality (as indicated by PESQ) and higher intelligibility (as shown by STOI). Second, DCUC-Net outperforms current and previous years' baselines. Finally, DCUC-Net with conformers for combining audio and visual information outperforms all other models (especially the one that used Bidirectional Long Short-Term Memory (BLSTM) and transformer for combining audio and visual information) in both evaluation metrics, providing strong evidence for the efficacy of integrating the conformer blocks into DCUC-Net model architecture.

\begin{table}[htb]
\centering
\caption{Objective assessment scores of noisy speech and enhanced speech generated by the baseline and the proposed DCUC-net on the COG-MHEAR dataset.}
\begin{tabular}[c]{lll}
    \hline
      & PESQ & STOI \\
    \hline
    Noisy & 1.15 & 0.64 \\
    Baseline (2022)\cite{challenge2023} & 1.30 & 0.67 \\
    Baseline (2023)\cite{challenge2023} & 1.70 & 0.83 \\
    Ours (BLSTM) & 1.62 & 0.81 \\
    Ours (Transformer) & 1.79 & 0.83 \\
    Ours (Conformer) & {\bf 1.84} & {\bf 0.84} \\
    \hline
\end{tabular}
\end{table}

\begin{figure}[htb]
    \begin{tabular}{cc}
        \includegraphics[width=4.7cm]{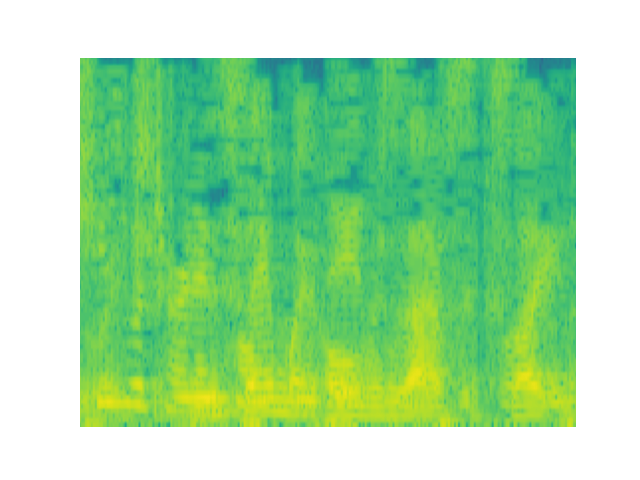} \hspace{-1cm} & \includegraphics[width=4.7cm]{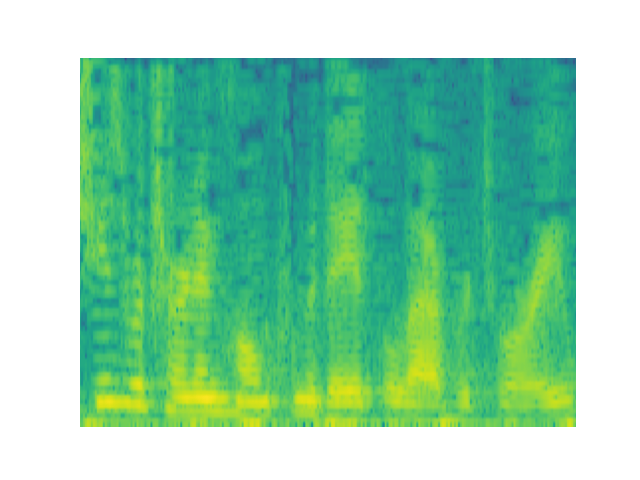} \\
        (a) Noisy spectrogram & (b) Clean spectrogram\\
        \includegraphics[width=4.7cm]{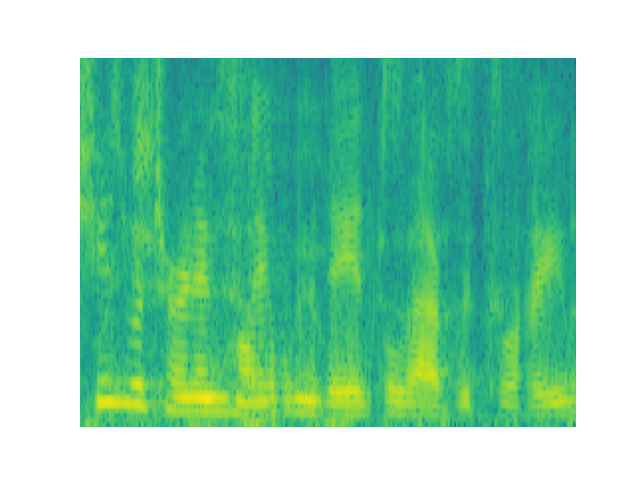} \hspace{-1cm} & \includegraphics[width=4.7cm]{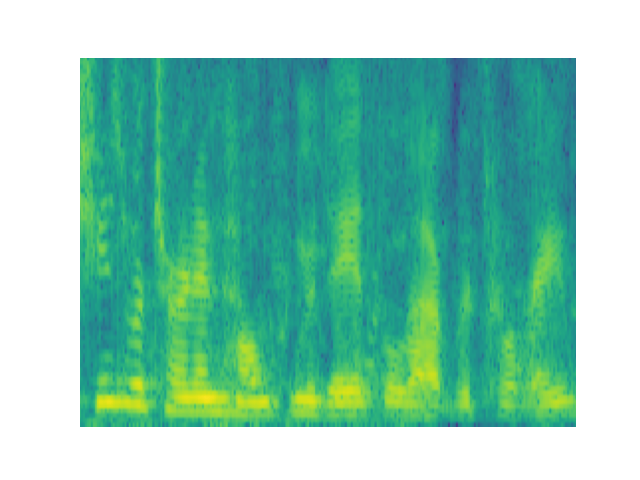} \\
        (c) Baseline AVSE & (d) Proposed DCUC-Net \\
    \end{tabular}
    \caption{Spectrograms of (a) noisy speech, (b) clean speech, (c) enhanced speech by the COG-MHEAR AVSE Challenge 2023 baseline, and (d) enhanced speech by DCUC-Net for an example in the COG-MHEAR development dataset. The vertical axis represents frequency, and the horizontal axis represents time.}
    \label{fig:grid}
\end{figure}

We performed spectrogram analysis on noisy speech, clean speech, and enhanced speech generated by the COG-MHEAR AVSE Challenge 2023 baseline and the proposed DCUC-Net. Figure 4 shows the corresponding spectrograms associated with an example in the development set. It is obvious from the figure that our model effectively suppresses the noise components present in the noisy speech spectrogram. Comparing the enhanced speech spectrograms of DCUC-Net and the baseline model, DCUC-Net provides significantly better noise reduction capabilities than the baseline model.

\subsection{Evaluation Results on the TMSV Dataset}
\label{ssec:TMSV results}

\begin{table}[htb]
\centering
\caption{Objective assessment scores of noisy speech and enhanced speech generated by various AVSE models on the TMSV dataset.}
\begin{tabular}[c]{lll}
    \hline
      & PESQ & STOI \\
    \hline
    Noisy & 1.18 & 0.60 \\
    LogMMSE(Audio-only)\cite{1495469} & 1.21 & 0.61 \\
    AVCVAE \cite{9110765} & 1.34 & 0.63 \\
    LAVSE \cite{chuang2022improved} & 1.31 & 0.61 \\
    SSL-AVSE \cite{10193049}  & 1.40 & {\bf 0.68} \\
    Ours (Conformer) & {\bf 1.41} & 0.66 \\
    \hline
\end{tabular}
\end{table}

We also evaluated noisy speech and enhanced speech generated by different SE and AVSE models on the TMSV dataset, including one audio-only traditional SE approach (MMSE), and three deep learning-based AVSE systems (AVCVAE \cite{9110765}, LAVSE \cite{chuang2022improved}, and SSL-AVSE \cite{10193049}). From Table 2, we can see that DCUC-Net outperforms all compared models in PESQ and STOI metrics, except SSL-AVSE \cite{10193049} in STOI metric. It is worth noting that AVSE adopts the pre-trained self-supervised learning model AV-Hubert for feature extraction, but the proposed DCUC-Net is trained from scratch. The model size of DCUC-Net is also much smaller than SSL-AVSE. Therefore, DCUC-Net has its advantages in terms of practical implementation. 

%our findings revealed that our model's performance is comparable to that of the state-of-the-art model for the TMSV dataset, which utilizes feature extractions from AV-Hubert as embeddings for AVSE. Detailed results of the model comparisons are presented in Table 2.

\section{CONCLUSIONs}
\label{sec:conclusion}
This paper introduces a novel AVSE framework in which we incorporate lip movement cues as visual features into the audio stream, and we emphasize the importance of phase information by leveraging complex features in a deep complex U-Net architecture. The visual features are combined with the output of the complex encoder for the audio stream, and the resulting concatenated features are processed by conformer blocks. The proposed framework effectively captures global and local audio-visual dependencies. Experimental results on the COG-MHEAR dataset demonstrate the superior performance of the proposed DCUC-Net AVSE framework over two baselines. Furthermore, on the TMSV dataset, DCUC-Net performs comparably to a state-of-the-art model that uses the pre-trained self-supervised learning AV-HuBERT model for feature extraction, and outperforms all other compared models. 
% The results demonstrate the strong denoising capabilities of our model even in challenging environments.

% References should be produced using the bibtex program from suitable
% BiBTeX files (here: strings, refs, manuals). The IEEEbib.bst bibliography
% style file from IEEE produces unsorted bibliography list.
% -------------------------------------------------------------------------
\bibliographystyle{IEEEbib}
\bibliography{strings,refs}

\end{document}